\documentstyle[12pt]{article}
\begin{document}
\begin{titlepage}
\pagestyle{empty}
%\begin{flushright} 
%{BROWN-HET-} \\
%{September 1995}
%\end{flushright} 
\vspace*{1.0cm}
 \begin{center} 
{\Large \bf Spectrum of $\gamma$-Fluids: A Statistical Derivation}
\\ [10mm] 
\renewcommand{\thefootnote}{\alph{footnote}} 
J. A. E. Carrillo$^{1,}$\footnote{e-mail:espichan@ifi.unicamp.br}\\ 
J. A. S. Lima$^{2,3,}$\footnote{e-mail:limajas@het.brown.edu}\\ A.
Maia Jr.$^{4,}$\footnote{e-mail:maia@ime.unicamp.br}
\end{center} 
\begin{flushleft}
\vspace{1.0cm}
1) {\it Instituto de F\'{\i}sica ``Gleb Wathagin" - UNICAMP
13.081-970 - Campinas - S.P. \hspace*{0.4cm}- Brazil.}\\          
2){\it Department of Physics, Brown University, Box 1843
Providence, RI 02912, USA.} \\
3){\it Departamento de F\'{\i}sica, UFRN, CP 1641
59072-970, Natal, Rio Grande do \hspace*{0.4cm}Norte, Brazil.}\\ 
4){\it Instituto de Matem\'atica - UNICAMP  
13.081-970 - Campinas - S.P. - Brazil.}\\
\end{flushleft} 

\begin{center}

\vspace{1.7cm} 

{\large \bf Abstract}
 
\end{center}

\vspace*{1.7cm}

The spectrum of massless bosonic and fermionic fluids satisfying the equation 
of state $p=(\gamma-1)\rho$ \ is derived using elementary statistical methods. 
As a limiting case, the Lorentz invariant spectrum of the vacuum 
($\gamma=0, p=-\rho$) is deduced. These results are in agreement with our 
earlier derivation for bosons using thermodynamics and semiclassical 
considerations.
\end{titlepage}

\baselineskip 0.65cm

\newpage

\noindent
{\large \bf 1. INTRODUCTION}

\vspace*{0.5cm}

\hspace{0.3in}The class of $\gamma$-fluids is the simplest kind of relativistic 
perfect simple fluids  used in the framework of general relativity and cosmology.
Such a class is usually defined in terms of so-called ``$\gamma$-law'' 
equation of state
\begin{eqnarray}
P=(\gamma -1)\rho  \quad,
\end{eqnarray}
where $\gamma \in [0,2]$. Some special types of media described by the
the above relation are: (i) vacuum ($p=-\rho $, $\gamma=0$) (ii)
a randomly oriented distribution of infinitely thin, straight
strings averaged over all directions ($p=-{1 \over 3} \rho $, $\gamma =2/3$), 
(iii) blackbody radiation ($p={1\over 3}\rho$, $\gamma = 4/3$), and 
(iv) stiff matter ($p=\rho$, $\gamma=2$). In a series
of recent papers (Lima and Santos, 1995; Lima and Maia, 1995a,b), some
general properties  of this monoparametric family of fluids have been discussed 
based on thermodynamic and 
semiclassical considerations. In particular, we have stressed the unusual 
thermodynamic behavior arising when the $\gamma$ parameter is smaller than
one. In this case, unlike of the subset with positive pressure, the 
temperature increases in the course of an adiabatic expansion. In the vacuum 
case, for instance, it was shown that the temperature scales as $T \sim V$, 
where V is the volume. Further, by assuming that such fluids may be regarded 
as a kind of generalized radiation, the general Planck form of the 
spectrum have been obtained, which includes the vacuum spectrum as a particular 
case (Lima and Maia, 1995b). In our opinion,
the special attention given to this class of fluids has a very simple
physical motivation. Physicists  have no intuitive
picture of the relativistic quantum vacuum, which remains  one the most 
unknown physical systems. A possible way to overcome such a difficulty is 
using a $\gamma$-fluid. In principle, by establishing the physical 
properties for a generic value of $\gamma$, one can obtain the Lorentz-invariant
vacuum properties taking the limit $\gamma=0$. Such a possibility, 
may be important even in the cosmological domain, where the vacuum physics 
is closely related to the cosmological constant problem (Weinberg, 1989). 
In this connection, it should be 
recalled that in some stages, the scalar fields driving inflation can also be 
thought of as a kind of $\gamma$-fluid, regardless of the details of its 
potential.
This happens, for instance, during the coherent field oscillations phase of
the inflaton field at the end of inflation (Kolb and Turner, 1990). 

In this context, it seems interesting to extend 
the classical thermodynamic approach developed in the above papers,
making the necessary connection with
the microphysics underlying such systems. In the present article, our main 
goal is to show how the
Planckian-type distribution for a $\gamma$-fluid, which has been discussed 
in the framework of the old quantum theory of radiation, can be 
reproduced in the domain of statistical mechanics. This allow us to extend
the theory for fermions as well. Of course, 
the third and last step would be to derive the spectrum from a more 
basic theory as quantum field theory.

\vspace*{0.8cm}

\noindent
{\large \bf 2. THE SPECTRUM OF $\gamma$-FLUIDS}

\vspace*{0.5cm}

\hspace{0.3in} Now consider the canonical procedure to compute the pressure
$p$ and the energy density $\rho$ in elementary statistical mechanics. 
As usually, these quantities are defined by 
\begin{eqnarray}
p=kT{\left({\partial \ell n Q \over \partial V}\right)}_{T}  \quad 
\label{rg2}
\end{eqnarray}
and
\begin{eqnarray}
\rho={kT^{2} \over V}{\left({\partial \ell n Q \over \partial T}\right)}_{V}
\label{rg92}  \quad
\end{eqnarray}
where $\ell n Q$ is the grand-canonical thermodynamic potential, which 
corresponds to a quantum fluid in contact with a thermal reservoir at 
temperature T. Since we are 
assuming that the vacuum state behaves like a kind of radiation, which 
differs  from blackbody radiation
only due to the equation of state, we take  the chemical potential 
of any $\gamma$-fluid to be identically zero. In this case, by considering 
a continuous spectrum, we have the well known formula (Itzykson and Zuber, 1980)
\begin{eqnarray} 
\ell n Q =-V\int_{0}^{\infty}\ell n \left(1\mp \exp(-\frac {\hbar\omega}{kT})
\right)f(\omega)d\omega \quad 
\label{rg4}
\end{eqnarray} 
where the upper and lower sign inside the brackets correponds 
to bosons and fermions respectively.

Our aim now is to find the unknown function $f(\omega)$, which is the number 
of states per unit energy.

From equations (1)-(4) we get easily
\begin{eqnarray}
-kT\int_{0}^{\infty}\ell n \left(1\mp \exp(-{\hbar\omega\over kT})\right)
f(\omega)d\omega=(\gamma -1)\hbar\int_{0}^{\infty}
{\omega f(\omega)\over \exp({\hbar\omega \over kT})\mp 1}d\omega. 
\label{rg5}
\end{eqnarray}

Of course, the above equation points to a singularity at $\gamma=1$. 
This rather pathological 
case (``dust''), describing a zero-pressure fluid, will not be 
considered here. A partial integration on the left-hand side of (\ref{rg5}) 
furnishes
\begin{eqnarray}
-kT\ell n \left(1\mp \exp(-{\hbar\omega\over kT})\right)F(\omega) \biggr]_{0}^{\infty}+\hbar\int_{0}^{\infty}{F(\omega)\over \exp({\hbar\omega\over kT})\mp 1}d\omega,
\label{rg6}
\end{eqnarray}
where $F(\omega)$ is a primitive of $f(\omega)$
\begin{eqnarray}
{F}^{\prime}(\omega)=f(\omega).
\label{rg7}
\end{eqnarray}

Let us now suppose, for a moment, that the first term in (\ref{rg6}), which 
corresponds to a boundary term, vanishes. In what follows, it will become clear 
under which conditions the function 
$f(\omega)$ will fulfill such a constraint. Bearing this in mind,
we may write from (5) and (6)

\begin{eqnarray}
\int_{0}^{\infty}{F(\omega)\over \exp({\hbar\omega\over kT})\mp 1}
d\omega=(\gamma -1)\int_{0}^{\infty}{\omega f(\omega)\over 
\exp({\hbar\omega \over kT})\mp 1}d\omega.
\label{rg8}
\end{eqnarray}

The correctness of the above equation will be guaranteed if the functions 
$f(\omega)$ and $F(\omega)$ obeying (7) satisfy the following relation
\begin{eqnarray}
F(\omega)=(\gamma -1)\omega f(\omega).
\label{rg9}
\end{eqnarray}
In principle, we cannot guarantee that equation (\ref{rg9}) will furnish
all physically meaningful solutions of  equations (7) and (8). Our confidence 
that it is the physical solution is supported by our equivalent earlier 
result using only
thermodynamics and semiclassical considerations (Lima and Maia, 1995b). In 
addition, 
it is easy to see that equation (\ref{rg9}) is independent of the statistics 
of the $\gamma$-fluids particles.

From equations (\ref{rg7}) and (\ref{rg9}) one obtains the differential 
equation for $f(\omega)$

\begin{eqnarray}
{f^{\prime}(\omega) \over f(\omega)}=({2- \gamma \over \gamma - 1}){1 \over \omega} \quad ,
\label{rg10}
\end{eqnarray}
where the prime denotes derivation with respect to $w$. The solution of above
equation is straightforward,
\begin{eqnarray}
f(\omega)=A {\omega}^{(2- \gamma)/(\gamma -1)} \quad ,
\label{rg11}
\end{eqnarray}
where A is a $\gamma$-dependent integration constant. Now, inserting the above 
equation into (4) and using (3) we obtain
\begin{eqnarray}
\rho(T)= \int_{0}^{\infty}{A\omega^{1\over \gamma -1} \over \exp({\hbar\omega\over kT})\mp 1}d\omega.     
\end{eqnarray}

Therefore, the spectrum of a $\gamma$-fluid reads:
\begin{eqnarray}
\rho(\omega, T) = {A\omega^{1\over \gamma -1} \over exp({\hbar\omega\over kT})
\mp 1}.
\label{u13}
\end{eqnarray}

For the case of bosons, equations (12) and (13) above are exactly equations 
(39) and (53) presented by Lima and Maia(1995b). As expected, by introducing a 
new variable $x=\frac{\hbar\omega}{kT}$, one obtains from (12), 
the generalized Stefan--Boltzmann law (Lima and Santos, 1995)
\begin{equation}
\label{ROTETA}
\rho (T)=\eta T^{{\gamma \over \gamma - 1}},
\end{equation}
where the constant $\eta$ depends on the $\gamma$-parameter as well as on the 
bosonic (or fermionic) spin degrees of freedom of each field. Note also that 
the above expression for $\rho(T)$ does not 
means that the energy density is always finite for any value of $\gamma$. In 
particular, for the vacuum case ($\gamma=0$), $\rho$ effectively does not 
depend on the temperature, but the constant $\eta$ is infinite, as it should 
be from quantum field theory.

\vspace*{0.8cm}

\noindent
{\large \bf 3. THE VACUUM-INFRARED DIVERGENCE}

\vspace*{0.5cm}

\hspace{0.3in} Naturally, the validity of equations (12)-(14), is crucially 
dependent on our earlier hypotheses concerning the boundary term in equation 
(6). In order to clarify this point we will compute explicitly such a term. 
From (9) and (11) it follows that
\begin{eqnarray}
F(\omega)=A(\gamma -1){\omega}^{{1 \over \gamma-1}} \quad.
\label{rg12}
\end{eqnarray} 
Now, inserting the function $F(\omega)$ into (6), we see that
for $\gamma$ ranging on the interval $1 < \gamma\leq 2$,
the boundary term vanishes in accordance with our earlier conjecture. In 
particular, this means that the above derivation works well in 
the case of photons $(\gamma=4/3)$. However, we find a 
divergence in the limit $\omega \rightarrow 0$, when $0 \leq \gamma < 1$.  
In the vacuum case, for instance, equation (13) reduces to
\begin{eqnarray}
{\rho}_{vac}(\omega,T)={A \hbar {\omega}^{-1} \over 
\exp({\hbar \omega \over kT})\mp 1}.
\end{eqnarray}

Thus, even though that the vacuum 
energy density does not depends on the temperature [see (15)],
it also exhibts the same kind of divergence.
Thus the spectrum for negative pressures $0\leq\gamma <1$, demands  
closer attention due to the inevitable existence of an infrared 
divergence.

To avoid the infrared catastrophe we proceed in analogy with the Casimir effect,
in which the divergent energy density has been regularized by an ultraviolet 
exponential cut-off $e^{-\alpha\omega}$, with $\alpha >0$ (Plunien et al., 1986; 
Ruggiero and Zimmerman, 1977). 
In this way, we use an infrared exponential
cut-off $e^{-{\alpha\over\omega}}$, $\alpha >0$. By introducing  
the regularized function
\begin{eqnarray}
F_{\alpha}(\omega)=F(\omega)e^{-{\alpha\over\omega}} \quad ,
\end{eqnarray}
it is straighforward to check that $F_{\alpha}(\omega)$ makes the boundary
term in (\ref{rg6}) vanishes.

Now, returning to equation (\ref{rg7}), we may 
define its regularized counterpart
\begin{eqnarray}
f_{\alpha}(\omega)=F^{\prime}_{\alpha}(\omega)
\end{eqnarray}
and from (15), (17) and (18) is readily  obtained the regularized density of 
states function
\begin{eqnarray}
f_{\alpha}(\omega)=A(1+{(\gamma -1)\alpha \over \omega})
{\omega}^{(2-\gamma)/(\gamma - 1)}\exp(-{\alpha \over \omega}) \quad ,
\label{rg27}
\end{eqnarray}
which, as should be expected, reduces to $f(\omega)$ in the limit 
$\alpha \rightarrow 0$. 
From (5) and (6) the regularized equation of state reads
\begin{eqnarray}
P_{\alpha}=(\gamma -1)\rho_{\alpha} \quad ,
\label{rg29}
\end{eqnarray}
where
\begin{eqnarray}
P_{\alpha}=\int_{0}^{\infty}{F_{\alpha}(\omega)\over \exp({\hbar\omega\over kT})\mp 1}d\omega 
\end{eqnarray}
and
\begin{eqnarray}
\rho_{\alpha}=\int_{0}^{\infty}{\omega f_{\alpha}(\omega)\over \exp^({\hbar\omega\over kT})\mp 1}d\omega.    
\end{eqnarray}

It should be noticed taht the above regularized integrals are finite to all 
cases $0\leq\gamma\leq 2, \gamma\neq 1$.
However, if $\alpha \rightarrow 0$, we obtain the original infrared divergence 
for $0\leq\gamma <1$. In particular, this means that the 
method outlined in section 2 is valid either
with no regularization or renormalization only for positive pressures.
As a matter of fact, although the regularized quantities $P_{\alpha}$ and
$\rho_{\alpha}$ are finite, they are cut-off dependent. To eliminate this
dependence, a renormalization scheme is required. In this connection it seems 
interesting to investigate the  
second quantization of $\gamma$-fluids and search for a renormalization 
scheme in this theory. This issue is presently under investigation.

\vspace*{0.8cm}

\noindent
{\large \bf REFERENCES}

\vspace*{0.5cm}

\noindent
Itzykson, C. and Zuber, J. B. (1980). 
{\it Quantum Field Theory}, McGraw-Hill, New York.

\noindent
Kolb, E. W. and Turner, M. S. (1990). 
{\it The Early Universe}, Addison-Wesley, Reading, Massachusetts.

\noindent
Lima, J. A. S. and Maia, A. Jr. (1995a). 
{\it International Journal of theoretical Physics}, {\bf 34}, 1835. 

\noindent 
Lima, J. A. S. and Maia, A. Jr.(1995b).
 {\it Physical Review D }. {\bf 52}, 5628.

\noindent
Lima, J. A. S. and Santos, J. (1995). {\it International Journal 
of Theorethical Physics}, {\bf 34}, 127.

\noindent
Plunien, G., M\"uller, B. and Greiner, W. (1986).
{\it Physics Reports}, {\bf 134}, 88.

\noindent
Ruggiero, J. R. and Zimmerman, A. H. (1977).
{\it Revista Brasileira de F\'{i}sica}, {\bf 7}, 663.

\noindent
Weinberg, S. (1989). {\it Review of Modern Physics}, 61, 1.

\end{document}